\documentclass[10pt,conference,letterpaper]{IEEEtran}

\usepackage[T1]{fontenc}
\usepackage[cmex10]{amsmath}
\usepackage{amssymb}
\usepackage{amsthm}
\usepackage{amsfonts}
\usepackage{algorithmic}
\usepackage{cite}
\usepackage{textcomp}
\IEEEoverridecommandlockouts
\usepackage{times}
\usepackage[scriptsize]{subfigure}
\usepackage{paralist}
\usepackage{mathrsfs}
\usepackage{graphicx}
\usepackage{tabularx}
\usepackage{url}
\usepackage{mathbbol}
\usepackage{booktabs}
\usepackage{verbatim}

\usepackage{bm}            
\newtheorem{theorem}{Theorem}
\newtheorem{definition}{Definition}
\newtheorem{lemma}[theorem]{Lemma}

\usepackage{xcolor}
\usepackage{multirow}
\usepackage{cuted}
\usepackage{stfloats}

\begin{document}

\title{Modeling and Analysis of mMTC Traffic in 5G Base Stations}

\author{
\IEEEauthorblockN{Fidan Mehmeti}
\IEEEauthorblockA{Department of Computer Science and Engineering\\
The Pennsylvania State University, USA\\
Email: fzm82@psu.edu}
\and
\IEEEauthorblockN{Thomas F. La Porta}
\IEEEauthorblockA{Department of Computer Science and Engineering\\
The Pennsylvania State University, USA\\
Email: tlp@psu.edu}}

\maketitle

\begin{abstract}
Massive Machine-Type Communications (mMTC) are one of the three
types of services that should be supported by 5G networks. These are distinguished by the need to serve a large number of devices which are characterized by non-intensive traffic and low energy 
consumption. While the sporadic nature of the mMTC traffic does not pose an exertion to efficient network operation, multiplexing the traffic from a large number of these devices within the cell certainly does. Therefore, planning carefully the network resources for this traffic is of paramount importance. 
To do this, the statistics of the traffic pattern that arrives at the base station should be known. To this end, in this paper, we derive the distribution of the inter-arrival times of the traffic at the base station from a general number of mMTC users within the cell, assuming a generic distribution of the traffic pattern by individual users. We validate our results on traces. Results show that 
adding more mMTC users in the cell 
increases the variability of the traffic pattern at the base station almost linearly, which is not the case with increasing the traffic generation rates. 
\end{abstract}

\begin{IEEEkeywords}
Traffic characteristics, 5G, mMTC. 
\end{IEEEkeywords}

\IEEEpeerreviewmaketitle

\section{Introduction}
\label{sec:intro}

5G services that are of mMTC type require support for a very high density of devices and low energy consumption~\cite{Wang2020}.
mMTC is the service category responsible for providing access to a large number of machine-type devices, such as those for sensing, monitoring, metering, and all other Internet-of-Things (IoT) related applications/services. Besides providing access to a large number of devices for this service, the other goal for these devices is to operate on an energy-efficient basis. As opposed to Ultra-Reliable Low-Latency Communications (URLLC), mMTC services are less stringent in terms of the network conditions needed for their successful operation, i.e., their service requirements are considerably more relaxed. 

However, while the traffic for a single device underlying mMTC service is sporadic with only a few packets transmitted at a time, given the expected density of IoT devices, 
multiplexing too many devices within the service area of a given cell 
may put the cell
at risk of collapse by pushing its operation towards the capacity limits. 
This imposes the need for careful resource and network planning in order to deliver the information of these devices in a timely manner. 

There are some important research questions that arise from the problem of network planning for mMTC traffic:
\begin{itemize}
\item Given the traffic intensity of a user and its channel conditions, what is the amount of resources that are needed for  a (generally large number) of mMTC devices to send their data \emph{successfully} and \emph{timely} without collapsing the cellular network?
\item Given the finite network capacity, is it possible to serve a very large number of mMTC users within the cell without forcing the network to operate close to its critical capacity limits? 
\end{itemize}

To answer these questions, the knowledge of \emph{traffic patterns} in the network is of paramount importance. With knowledge of traffic characteristics, resource planning can be conducted successfully. Also, given the limited network resources, deriving the traffic patterns can help the cellular operator to  determine the maximum number of mMTC users that can be served in a cell.   

In this paper, we characterize completely the mMTC traffic pattern in the cell by deriving the distribution of inter-arrival times of the mMTC traffic at the base station and the distribution of the number of packets, given the statistics of traffic generation at individual mMTC users.  
The analysis we provide in this paper relies on realistic assumptions, such as general distributions for the inter-generation times of data at mMTC users, and general distributions for the number of packets generated at once. 
The results that we provide can help cellular network operators in planning the number of users that can be served (admission control), or in planning the network resources needed to provide service to a given number of mMTC users.    
\textcolor{black}{The main message of this paper is that the traffic pattern that mMTC users generate and their number in the cell are far more decisive on the network traffic pattern than the channel conditions of mMTC users.} 

Specifically, our main contributions are:
\begin{itemize}
\item We derive the distribution of inter-arrival times at the base station that holds for any distribution of the inter-generation times of the traffic at mMTC.  
\item We also derive the distribution of the number of packets that arrive at once at the base station, for any distribution of the number of packets generated at once on mMTC devices within the cell. 
\item Using extensive realistic simulations, conducted on real-life traces, \textcolor{black}{we show that the coefficient of variation of inter-arrival times at gNodeB is oblivious to the  channel conditions and traffic generation rates at mMTC users.} 
\end{itemize}

The remainder of this paper is organized as follows. In Section~\ref{sec:model}, we introduce the system model, including the mMTC traffic. This is followed by the analysis for the distribution of inter-arrival times and the number of packets 
in Section~\ref{sec:analysis}.   
Then, we perform an evaluation on real traces in Section~\ref{sec:sims}.  
In Section~\ref{sec:related}, we discuss some related work. Finally, Section~\ref{sec:conclusion} concludes the paper. 

\section{Performance Modeling}
\label{sec:model}
In this section, we present the system model and traffic characteristics of mMTC users. 

\subsection{System model}
We consider users within the coverage area of a \textcolor{black}{5G} macro base station (\textcolor{black}{gNodeB}) \textcolor{black}{in the} sub-6 GHz band. Due to the nature of mMTC traffic, we focus on the uplink. 

The block resource allocation scheme in a time-slotted system (frames) is used in 5G, with \emph{physical resource blocks (PRB)} being the allocation unit~\cite{block}.  
Within the frame, different blocks are assigned to different users. In general, the assignment will vary across frames.  As a result, scheduling needs to be performed along two dimensions, \emph{frequency} and \emph{time}. 
The total number of available blocks for the users with mMTC traffic in the cell is $K$. 

\textcolor{black}{
The possibility of network slicing in 5G~\cite{p11} enables assigning \emph{dedicated} network resources to the same type of service, e.g., mMTC users.
Therefore, throughout this paper, we assume that all mMTC users belong to the same slice within the cell, and hence require the same service quality.}  


In general, users will see different channel conditions in different frequencies (different blocks) even within the same frame. Therefore, they  will experience a different per-block \textcolor{black}{Signal-to-Interference-plus-Noise Ratio} (SINR). This  \textcolor{black}{is a function of the base station transmission power of the cell where users are, the transmission power of neighboring cells transmitting on the same frequency set (inter-cell interference), the background noise, and the corresponding channel gains~\cite{Fidan_TMC}.} Because of 
the time-varying channel characteristics, the  per-block SINR changes from one frame to another even for the same block for every user. This changing \emph{per-block} SINR translates into a varying \emph{per-block rate} that is obtained using a modulation and coding scheme (MCS). In our system, we consider the MCS with $m$ possible values (the typical value of $m$ is 15)~\cite{3GPP_5G_NR}. 

Further, for every user, we assume flat channels \textcolor{black}{(blocks)} in a frame, i.e., the per-block rate does not change during the frame, but it changes from one frame to another randomly. 

Although, in general, different blocks will ``bring'' different per-block rates, for the sake of analytical tractability, 
we make a simplifying assumption. Namely, 
we assume that the gNodeB transmission power and channel characteristics of a user remain unchanged across all $K$ blocks in a frame. i.e., they undergo flat fading. 
Consequently, our problem reduces to one-dimensional scheduling, in \emph{time}. Therefore, instead of deciding how many and which blocks to assign to every user, we use another parameter, defined as:
\begin{definition}
The ratio of the frame for which all the blocks are allocated to user $i$ is called \textbf{frame ratio}. It is denoted by $Y_i$ and can take values in the interval $[0,1]$. 
\end{definition}

The blocks are assigned orthogonally during the frame so that no two  users receive them simultaneously. 
\textcolor{black}{This is a reasonable simplification as frame ratio can be translated into the corresponding number of PRBs.}
Hence, the frame ratio will be the quantitative measure of resource allocation in this paper. 

The previous assumptions enable us to model user's $i$ per-block rate as a discrete random variable, $R_{i}$, with values in $\left\{r_{1},r_{2},\ldots,r_{m}\right\}$, such that $r_{1}<r_{2}<\ldots<r_{m}$, \textcolor{black}{with a probability mass function (pmf) $p_{R_{i}}(x)$. The latter is a function of  user's $i$ SINR over time.}

\emph{Number of users:} There are $n$ users in the cell with different per-block rate distributions.


\subsection{mMTC traffic}
As already mentioned, we consider the communication on the uplink. The data from devices are sent to the base station that covers the cell where devices are located. 

\emph{Traffic generation:} Data from every user to the base station are transmitted in packets. We assume that packet sizes, denoted as $\Theta$, are fixed~\cite{Ortiz}. This is reasonable as we are considering services in which the data are organized in small packets~\cite{Ortiz}. Data are generated with rate $\lambda$, which is called \emph{traffic generation rate}, and which represents the inverse of the average time between two consecutive transmissions of data from mMTC user. The mMTC traffic is sporadic, which means that it is characterized by small $\lambda$. The time interval between two consecutive data generation instants, which from now on we will be referring to as \emph{inter-generation time}, undergoes a general distribution.     

\emph{Number of packets:} At the moment of generation, we assume that there are several packets that are transmitted. The random variable $N$ is used to denote the number of packets that are generated at a given time. We assume that $N(t)$ is a stationary stochastic process that is independent of the data rate (channel conditions) of the user.\textcolor{black}{\footnote{We remove the reference to $t$ for notational simplicity from now on.}} The maximum number of packets that an mMTC user can generate at once is $N_{max}$.   

\textcolor{black}{Users operate independently from one another, i.e., the data generation instants and the number of packets at a specific user aren't in any way correlated with data generation instants and the number of packets of other users.}

\emph{Data rate:} The data rate a user experiences in a frame is $U_i=KY_iR_i$, where $Y_i$ is the frame ratio, $K$ is the total number of blocks, and $R_i$ is the random variable that denotes the per-block rate of user $i$ in a frame. In this paper, we assume that network resources are shared equally among the users, which implies a data rate of $U_i=\frac{KR_i}{n}$ for user $i$, where $n$ is the number of mMTC users in the cell. 

As packet sizes are small, and not too many packets are transmitted at once, we assume that all the packets generated at once are transmitted with the same rate $U$. 

\section{Analysis}
\label{sec:analysis}
In this section, we first derive the distribution of inter-arrival times of the traffic at gNodeB. Then, we derive the distribution of the number of packets arriving at gNodeB. 

\subsection{Distribution of inter-arrival times}
First, we derive the distribution of inter-arrival times at the gNodeB from a single user. Then, we use this result to derive the distribution of inter-arrival times at gNode from all mMTC users in the cell. 
\subsubsection{Single-user analysis}
\label{sec:analysis_single}
Let $t_i$ denote the time when the $i$'th amount of data are generated and  transmitted from a user to the base station. The number of packets in that batch is $N_i$. Note that $t_0=0$ and $N_0=0$. The time all these packets arrive at the base station is denoted by $t_i^{'}$, and is
\begin{equation}
t_i^{'}=t_i+\frac{N_i\Theta}{U_i}+t_{prop},
\label{eq:time_i}
\end {equation}
where $\frac{N_i\Theta}{U_i}$ represents the transmission delay, and $t_{prop}$ is the propagation delay from the user to the base station. Note that since we assume the mMTC traffic is sporadic and packet sizes are small, it is reasonable to assume that all the packets generated in an instant are to be transmitted within the same frame, whose duration is $10$\:ms~\cite{3GPP_5G_NR}.

The time instant at which the next batch of data from the same user arrives at the base station is
\begin{equation}
t_{i+1}^{'}=t_{i+1}+\frac{N_{i+1}\Theta}{U_{i+1}}+t_{prop}.
\label{eq:time_i+1}
\end{equation}
From Eqs.(\ref{eq:time_i}) and (\ref{eq:time_i+1}), we obtain the inter-arrival time between two data arrivals at gNodeB from the same user as
\begin{equation}
\Delta t^{'}=t_{i+1}^{'}-t_{i}^{'}=\Delta t + \Theta\left(\frac{N_{i+1}}{U_{i+1}}-\frac{N_i}{U_i}\right), 
\label{eq:t_'}
\end{equation}
where $\Delta t= t_{i+1}-t_i$ is the period between two consecutive data generations at the mMTC user. Since we are assuming the equal-share policy for the network resources, the data rate of the user when generating the $i$'th batch of data is $U_i=\frac{KR_{i}}{n}$, and $U_{i+1}=\frac{KR_{i+1}}{n}$ when generating the $i+1$'th batch of data. Introducing the last assumptions into Eq.(\ref{eq:t_'}), we get
\begin{equation}
\Delta t^{'}=\Delta t + \frac{n\Theta}{K}\left(\frac{N_{i+1}}{R_{i+1}}-\frac{N_{i}}{R_{i}}\right). 
\label{eq:t_'_updated}
\end{equation}
As the number of generated packets and user's per-block rate are independent from the moment when data are generated ($N(t)$ is a stationary stochastic process), $\Delta t$ is independent from the term $\frac{n\Theta}{K}\left(\frac{N_{i+1}}{R_{i+1}}-\frac{N_{i}}{R_{i}}\right)$ 
in Eq.(\ref{eq:t_'_updated}). Therefore, the probability density function (pdf) of $\Delta t^{'}$ can be written as~\cite{RossStochProc} 
\begin{equation}
f_{\Delta t^{'}}(x)=f_{\Delta t}(x) * f_{\frac{n\Theta}{K}\left(\frac{N_{i+1}}{R_{i+1}}-\frac{N_i}{R_i}\right)}(x),  
\label{eq:f}
\end{equation}
where * denotes the convolution operation~\cite{Oppenheim}. 
\textcolor{black}{In Eq.(\ref{eq:f}), the second right-hand side term is a probability mass function (pmf). Namely, both $N$ and $R$ are discrete random variables (with values from finite discrete sets). Therefore, any combination of these two random variables with a constant ($\frac{n\Theta}{K}$) will take values from a finite discrete set. Hence, $\frac{n\Theta}{K}\left(\frac{N_{i+1}}{R_{i+1}}-\frac{N_i}{R_i}\right)$ is a discrete random variable, and $f_{\frac{n\Theta}{K}\left(\frac{N_{i+1}}{R_{i+1}}-\frac{N_i}{R_i}\right)}(x)$ is its pmf.\footnote{The inter-generation time of data at an mMTC user is a continuous random variable, the same as the inter-arrival time at gNodeB. Hence, the use of \emph{probability density function} (pdf) for the inter-arrival time.}} 

Before proceeding with the derivation of the aforementioned pmf, let us show what happens to the pmf of a function when the latter is multiplied by a constant - in this case $\frac{n\Theta}{K}$. 
\begin{lemma}
For random variable $X$ and constants $n, \Theta, K$, the following holds:
\begin{equation}
f_{\frac{n\Theta X}{K}}(x)=f_{X}\left(\frac{Kx}{n\Theta}\right).   
\label{eq:lemma_1}
\end{equation}
\end{lemma}
\begin{proof}
$$f_{\frac{n\Theta X}{K}}(x)=\mathbb{P}\left(\frac{n\Theta X}{K}=x\right)=\mathbb{P}\left(X=\frac{Kx}{n\Theta}\right)=f_{X}\left(\frac{Kx}{n\Theta}\right)$$
\end{proof}
\noindent In order to proceed with the derivation, we need the following:
\begin{lemma}
The pmf of the difference between the ratio $\frac{N}{R}$ in two consecutive generations of data is 
\begin{align}
f_{\frac{N_{i+1}}{R_{i+1}}-\frac{N_i}{R_i}}(x)&=\sum_{i=1}^{m}\sum_{k=1}^{N_{max}}\sum_{j=1}^{m}\sum_{l=1}^{N_{max}}p_N(k)p_R(r_i)p_N(l)p_R(r_j)\nonumber\\
&\cdot\delta\left(x-\frac{k}{r_i}+\frac{l}{r_j}\right),
\label{eq:NRx_8}
\end{align}
where $\delta(x)$ is the Dirac delta function~\cite{Oppenheim}, whose value is $1$ only at $x=0$, and $0$ otherwise. 
\end{lemma}
\begin{proof}

The pmf of the random variable $\frac{N}{R}$ is
\begin{equation}
f_{\frac{N}{R}}(x)=\mathbb{P}\left(\frac{N}{R}=x\right)=\mathbb{P}\left(N=Rx\right)=f_{N}(Rx).     
\end{equation}
For the latter, we have
\begin{eqnarray}
\mathbb{P}(N=Rx)&=\sum_{i=1}^{m}\mathbb{P}\left(N=Rx|R=r_i\right)\mathbb{P}\left(R=r_i\right)\nonumber\\
&=\sum_{i=1}^{m}\mathbb{P}\left(N=r_ix\right)p_R(r_i),
\label{eq:NRx}
\end{eqnarray}
where $p_R(r_i)$ denotes the probability mass function (pmf) of the per-block rate of a user. For the other term on the RHS of Eq.(\ref{eq:NRx}), we have
\begin{equation}
\mathbb{P}\left(N=r_ix\right)=p_{N}\left(r_ix\right)=\sum_{k=1}^{N_{max}}p_{N}(k)\delta\left(x-\frac{k}{r_i}\right).
\label{eq:NRx_1}
\end{equation}
Note that $\sum_{k=1}^{N_{max}}p_{N}(k)=1$ and $N_{max}$ is the maximum number of packets that can be generated at once. 
Substituting Eq.(\ref{eq:NRx_1}) into Eq.(\ref{eq:NRx}), we obtain
\begin{equation}
f_{\frac{N}{R}}(x)=\sum_{i=1}^{m}\sum_{k=1}^{N_{max}}p_{N}(k)p_{R}(r_i)\delta\left(x-\frac{k}{r_i}\right).   
\label{eq:NRx_2}
\end{equation}
Next, we need to look at the difference of two i.i.d. random variables
$\frac{N}{R}-\frac{N}{R}=\frac{N}{R}+\left(-\frac{N}{R}\right)$.\footnote{Note that this is the difference of two \emph{random} variables, and not the difference of two ordinary variables. Hence, it is not $0$ in general.} So, we need the pmf of the random variable $-\frac{N}{R}$. It is
\begin{equation}
f_{\hspace{-3pt}-\frac{N}{R}}(x)\hspace{-3pt}=\mathbb{P}\left(-\frac{N}{R}=x\right)=\mathbb{P}\left(\frac{N}{R}=-x\right)= f_{\frac{N}{R}}(\hspace{-3pt}-x),   
\end{equation}
which replaced into Eq.(\ref{eq:NRx_2}) yields
\begin{equation}
f_{-\frac{N}{R}}(x)=\sum_{i=1}^{m}\sum_{k=1}^{N_{max}}p_{N}(k)p_{R}(r_i)\delta\left(-x-\frac{k}{r_i}\right).
\label{eq:NRx3}
\end{equation}
In line with Eqs.(\ref{eq:t_'}) and (\ref{eq:f}), the following holds
\begin{equation}
f_{\frac{N_{i+1}}{R_{i+1}}-\frac{N_i}{R_i}}(x)=f_{\frac{N}{R}}(x)*f_{\frac{N}{R}}(-x).   
\label{eq:NRx_4}
\end{equation}
Substituting Eqs.(\ref{eq:NRx_2}) and (\ref{eq:NRx3}) into Eq.(\ref{eq:NRx_4}) results in
\begin{eqnarray}
f_{\frac{N_{i+1}}{R_{i+1}}-\frac{N_i}{R_i}}(x)=\sum_{i=1}^{m}\sum_{k=1}^{N_{max}}p_N(k)p_R(r_i)\delta\left(x-\frac{k}{r_i}\right)\nonumber\\
*\sum_{j=1}^{m}\sum_{l=1}^{N_{max}}p_N(l)p_R(r_j)\delta\left(-x-\frac{l}{r_j}\right).
\label{eq:NRx_5}
\end{eqnarray}
Note that the convolution operation is characterized by the distributive property~\cite{Oppenheim}, i.e., $f(x)*\sum_{i} g_i(x)=\sum_i f(x)*g_i(x)$. Therefore, Eq.(\ref{eq:NRx_5}) transforms into 
\begin{align}
f_{\frac{N_{i+1}}{R_{i+1}}-\frac{N_i}{R_i}}(x)&=
\sum_{i=1}^{m}\sum_{k=1}^{N_{max}}\sum_{j=1}^{m}\sum_{l=1}^{N_{max}}p_N(k)p_R(r_i)p_N(l)p_R(r_j)\nonumber\\
&\cdot\delta\left(x-\frac{k}{r_i}\right)*\delta\left(-x-\frac{l}{r_j}\right). 
\label{eq:NRx_6}
\end{align}
For the convolution of two $\delta$ functions from Eq.(\ref{eq:NRx_6}), we have
\begin{align}
&\delta\left(x-\frac{k}{r_i}\right)*\delta\left(-x-\frac{l}{r_j}\right)=\delta\left(-x-\frac{l}{r_j}\right)*\delta\left(x-\frac{k}{r_i}\right)\nonumber\\
&=\delta\left(-\left(x-\frac{k}{r_i}\right)-\frac{l}{r_j}\right)=\delta\left(-x+\frac{k}{r_i}-\frac{l}{r_j}\right).
\label{eq:aux_0}
\end{align}
From the definition of the Dirac delta function,
$\delta\left(-x+\frac{k}{r_i}-\frac{l}{r_j}\right)=1$ 
only when 
$-x+\frac{k}{r_i}-\frac{l}{r_j}=0$, 
which yields
$x=\frac{k}{r_i}-\frac{l}{r_j}$. Therefore,
\begin{equation}
\delta\left(-x+\frac{k}{r_i}-\frac{l}{r_j}\right)=\delta\left(x-\frac{k}{r_i}+\frac{l}{r_j}\right). 
\label{eq:NRx_7}
\end{equation}
Substituting Eq.(\ref{eq:NRx_7}) into Eq.(\ref{eq:aux_0}), and the latter into Eq.(\ref{eq:NRx_6}), we obtain Eq.(\ref{eq:NRx_8}). 
\end{proof}

\noindent Using the results from Lemma~1 and Lemma~2, we obtain:
\begin{theorem}
The pdf of inter-arrival times between two consecutive data batches at gNodeB from a single mMTC user is
\begin{align}
f_{\Delta t^{'}}(x)&=\sum_{i=1}^{m}\sum_{k=1}^{N_{max}}\sum_{j=1}^{m}\sum_{l=1}^{N_{max}}p_N(k)p_R(r_i)p_N(l)p_R(r_j)\nonumber\\
&\cdot f_{\Delta t}\left(x-\frac{n\Theta}{K}\left(\frac{k}{r_i}-\frac{l}{r_j}\right)\right). 
\label{eq:NRx_13}
\end{align}
\end{theorem}
\begin{proof}
From Eq.(\ref{eq:lemma_1}), we have
\begin{equation}
f_{\frac{n\Theta}{K}\left(\frac{N_{i+1}}{R_{i+1}}-\frac{N_i}{R_i}\right)}(x) =f_{\frac{N_{i+1}}{R_{i+1}}-\frac{N_i}{R_i}}\left(\frac{Kx}{n\Theta}\right),
\label{eq:cx}
\end{equation}
which from Lemma~2 yields
\begin{align}
f_{\frac{n\Theta}{K}\left(\frac{N_{i+1}}{R_{i+1}}-\frac{N_i}{R_i}\right)}(x)&\hspace{-3pt}=\hspace{-3pt}\sum_{i=1}^{m}\sum_{k=1}^{N_{max}}\sum_{j=1}^{m}\sum_{l=1}^{N_{max}}\hspace{-3pt}p_N(k)p_R(r_i)p_N(l)p_R(r_j)\nonumber\\
&\cdot\delta\left(\frac{Kx}{n\Theta}-\frac{k}{r_i}+\frac{l}{r_j}\right).
\label{eq:NRx_9}
\end{align}
Following a similar reasoning for the argument of $\delta$ as before, we have 
\begin{equation}
\delta\left(\frac{Kx}{n\Theta}-\frac{k}{r_i}+\frac{l}{r_j}\right)=\delta\left(x-\frac{n\Theta}{K}\left(\frac{k}{r_i}-\frac{l}{r_j}\right)\right).
\label{eq:NRx_10}
\end{equation}
Substituting Eq.(\ref{eq:NRx_10}) into Eq.(\ref{eq:NRx_9}), we obtain
\begin{align}
f_{\frac{n\Theta}{K}\left(\frac{N_{i+1}}{R_{i+1}}-\frac{N_i}{R_i}\right)}(x)&=\sum_{i=1}^{m}\sum_{k=1}^{N_{max}}\sum_{j=1}^{m}\sum_{l=1}^{N_{max}}p_N(k)p_R(r_i)p_N(l)p_R(r_j)\nonumber\\
&\cdot\delta\left(x-\frac{n\Theta}{K}\left(\frac{k}{r_i}-\frac{l}{r_j}\right)\right).
\label{eq:NRx_11}
\end{align}
As for the convolution operation the Dirac delta function is the unit element and it holds that $f(x)*\delta(x-x_0)=f(x-x_0)$, we have
\begin{equation}
f_{\Delta t}(x)*\delta\left(x-\frac{n\Theta}{K}\left(\frac{k}{r_i}-\frac{l}{r_j}\right)\right)=f_{\Delta t}\left(x-\frac{n\Theta}{K}\left(\frac{k}{r_i}-\frac{l}{r_j}\right)\right)  
\label{eq:NRx_12}
\end{equation}
Substituting Eq.(\ref{eq:NRx_11}) into Eq.(\ref{eq:f}), and 
due to the distributive property of convolution and Eq.(\ref{eq:NRx_12}), we obtain Eq.(\ref{eq:NRx_13}).
\end{proof}
Eq.(\ref{eq:NRx_13}) represents the pdf of inter-arrival times of data at gNodeB from a specific user. Next, we derive the distribution of inter-arrival times to gNodeB from all the mMTC users in the cell.

\subsubsection{Multi-user analysis}
\label{sec:analysis_multiple}
Let $T_j$ denote the inter-arrival time of data at gNodeB from mMTC user $j\in\{1,\ldots,n\}$. When the process starts at $t=0$ the first data arrival is described by
\begin{equation}
    \min \{T_1,\ldots,T_n\}.
\end{equation}
After the first arrival and when the equilibrium is established, 
the stochastic arrival process at the gNodeB can be captured by (subscript $m$ specifies the multi-user case)
\begin{equation}
    \Delta t_m^{'}=\min \{T_1^{(e)},\ldots,T_n^{(e)}\},
\end{equation}
where $T_{j}^{(e)}$ denotes the \emph{excess distribution}\footnote{This is the time until the next arrival of data from user $j$ occurs.} of the inter-arrival time for packets from user $j$ at gNodeB.

The cumulative distribution function (CDF) of the inter-arrival time at the base station is
\begin{equation}
F_{\Delta t_{m}^{'}}(x)=\mathbb{P}(\Delta t_{m}^{'}\leq x)=\mathbb{P}\left(\min \{T_1^{(e)},\ldots,T_n^{(e)}\}\leq x\right).   
\label{eq:TB}
\end{equation}
If the minimum of a number of random variables is smaller than or equal to $x$, then each of these variables has to be smaller than or equal to $x$, leading to
\begin{equation}
\mathbb{P}(\min \{T_1^{(e)},\ldots,T_n^{(e)}\}\leq x)=\mathbb{P}(T_1^{(e)}\leq x, \ldots, T_n^{(e)}\leq x).        
\end{equation}
As the processes of data generation at the mMTC users are mutually independent, so are the inter-arrival times of their data at gNodeB. Hence, we have
\begin{equation}
\mathbb{P}(T_1^{(e)}\leq x, \ldots, T_n^{(e)}\leq x)=\prod_{j=1}^{n}\mathbb{P}(T_j^{(e)}\leq x)=\prod_{j=1}^{n}F_{T_j^{(e)}}(x)   \label{eq:min} 
\end{equation}
From the theory of \emph{renewal processes}\cite{RossStochProc}, it is known that the pdf of the excess distribution is given by
\begin{equation}
f_{T^{(e)}}(x)=\frac{1-F_{\Delta t^{'}}(x)}{\mathbb{E}[\Delta t ^{'}]}. 
\label{eq:Te}
\end{equation}
We have already shown in Section~\ref{sec:analysis_single} that the distribution of inter-arrival times from a user to gNodeB is not the same as the distribution of  inter-generation times for the data at the mMTC user. However, \emph{what happens with the average inter-arrival rate of data from an mMTC user to gNodeB}? The answer to this question is as follows.
\begin{lemma}
The inter-arrival rate of traffic at gNodeB from a specific user is identical to the inter-generation rate of the traffic of that same mMTC user, i.e., 
\begin{equation}
\lambda^{'}=\lambda.  
\label{eq:lambda=lambda'}
\end{equation}
\end{lemma}
\begin{proof}
The traffic generation rate at the user is 
\begin{equation}
\lambda=\frac{1}{\mathbb{E}[t_i-t_{i-1}]},
\label{eq:lambda}
\end{equation}
i.e., the inverse 
of the mean inter-generation times. 
On the other hand, the rate at which the traffic from the same user arrives at the base station is 
\begin{equation}
\lambda^{'}=\frac{1}{\mathbb{E}[t_i^{'}-t_{i-1}^{'}]}.
\label{eq:lambda'}    
\end{equation}
Substituting Eqs.(\ref{eq:time_i}) and (\ref{eq:time_i+1}) into Eq.(\ref{eq:lambda'}), we obtain
\begin{equation}
\lambda^{'}=\frac{1}{\mathbb{E}\left[\Delta t_i+\frac{n\Theta}{K}\left(\frac{N_{i}}{R(t_i)}-\frac{N_{i-1}}{R(t_{i-1})}\right)\right]},
\label{eq:lambda'_2}    
\end{equation}
where $\Delta t_i=t_i-t_{i-1}$ is the inter-generation time between the $i-1$'th and $i$'th data generation at the user.

Let us now take a closer look at the term $\mathbb{E}\left[\frac{N_{i}}{R(t_i)}-\frac{N_{i-1}}{R(t_{i-1})}\right]$. Since $N(t)$ is a stationary stochastic process independent of $R(t)$, and $R(t)$ is independent from one frame to another, 
we have
\begin{align}
\mathbb{E}\left[\frac{N_{i}}{R(t_i)}\right]&=\mathbb{E}\left[N_i\right]\mathbb{E}\left[\frac{1}{R(t_i)}\right]=\mathbb{E}\left[N_{i-1}\right]\mathbb{E}\left[\frac{1}{R(t_{i-1})}\right]\nonumber\\
=\mathbb{E}\left[\frac{N_{i-1}}{R(t_{i-1)}}\right],  
\end{align}
which after being substituted into Eq.(\ref{eq:lambda'_2}) implies
\begin{equation}
\lambda^{'}=\frac{1}{\mathbb{E}[\Delta t_i]}=\lambda.  
\end{equation}
\end{proof}
Next, for the CDF of inter-arrival times at gNodeB from all the mMTC users within the cell, we have:
\begin{theorem}
The CDF of inter-arrival times at gNodeB from $n$ mMTC users in the cell is 
\begin{equation}
F_{\Delta t_m^{'}}(x)=\prod_{i=1}^{n}\lambda_i\int_{0}^{x}\left(1-F_{\Delta t_i^{'}}(w)\right)dw, 
\label{eq:theorem_5}
\end{equation}
where $F_{\Delta t_{i}^{'}}(w)$ is the CDF of inter-arrival times of data from user $i$ to gNodeB. 
\end{theorem}
\begin{proof}
From Eq.(\ref{eq:Te}) and Eq.(\ref{eq:lambda=lambda'}), we get
\begin{equation}
f_{T_{i}^{(e)}}(x)
=\lambda_i\left(1-F_{\Delta t_{i}^{'}}(x)\right).  
\label{eq:Te_1}
\end{equation} 
The CDF of the excess distribution is 
\begin{equation}
F_{T_{i}^{(e)}}(x)=\int_{0}^{x}f_{T_{i}^{(e)}}(w)dw=\lambda_i\int_{0}^{x}(1-F_{\Delta t_{i}^{'}}(w))dw,
\label{eq:FTe}
\end{equation}
which substituted into Eq.(\ref{eq:min}) yields Eq.(\ref{eq:theorem_5}).
\end{proof}
Note that
\begin{equation}
F_{\Delta t^{'}}(w)=\int_{0}^{w}f_{\Delta t^{'}}(y)dy 
\label{eq:FTX_aux}
\end{equation}
is the CDF of inter-arrival times of data from an mMTC user to gNodeB. 

\subsection{Distribution of the number of packets}
Having derived the distribution of inter-arrival times of data at gNodeB, we proceed with looking into the distribution of data arriving at once during these arrival instants.

Determining the distribution of the number of packets at a time instant reduces to deciding which user the received data belong to. Therefore, the received data belong to user 1 (drawn from distribution $N_1$) with probability $p_1$, user 2 (drawn from distribution $N_2$), \ldots, user $n$ (drawn from distribution $N_n$). Hence, the pmf of the number of packets arriving at once at gNodeB is
\begin{equation}
p_N(x)=\sum_{i=1}^{n}p_ip_{N_i}(x).
\label{eq:pdf_N}
\end{equation}

The next step is deriving the probabilities $p_i, \forall i\in\{1,\ldots,n\}$. As mentioned previously, $p_i$ denotes the probability that the next arrival of data belongs to user $i$, i.e., the excess distribution of the inter-arrival time of user $i$ has the lowest value. According to this definition, for $p_i$ we have
\begin{equation}
p_i=\mathbb{P}\left(T_i^{(e)}\leq \min\{T_1^{(e)},\ldots,T_n^{(e)}\}\right), 
\label{eq:pi_0}
\end{equation}
where under the \emph{min} operator there is no $T_{i}^{(e)}$ term. The following holds:
\begin{theorem}
The pmf of the number of packets arriving at once at gNodeB is given by
\begin{equation}
p_{N}(x)=\sum_{i=1}^{n}p_{N_i}(x)\prod_{j=1, j\neq i}^{n}\int_{0}^{\infty}F_{T_{i}^{(e)}}(x)f_{T_j^{(e)}}(x)dx.  
\label{eq:N_final}
\end{equation}
\end{theorem}
\begin{proof}
Eq.(\ref{eq:pi_0}) is equivalent to
\begin{equation}
p_i=\mathbb{P}\left(T_i^{(e)}\leq T_1^{(e)},\ldots, T_i^{(e)}\leq T_n^{(e)}\right),    
\end{equation}
and further, as inter-arrival times from individual mMTC users to gNodeB are mutually independent and hence the independence of their excess distributions, we have
\begin{equation}
p_i=\prod_{j=1, j\neq i}^{n}\mathbb{P}\left(T_{i}^{(e)}\leq T_{j}^{(e)}\right). \label{eq:pi}   
\end{equation}
Next, we need to derive $\mathbb{P}\left(T_{i}^{(e)}\leq T_{j}^{(e)}\right)$. We do this by conditioning on the value of $T_{j}^{(e)}$. This yields
\begin{equation*}
\mathbb{P}\left(T_{i}^{(e)}\leq T_{j}^{(e)}\right)=\int_{0}^{\infty}\mathbb{P}\left(T_{i}^{(e)}\leq T_{j}^{(e)}|T_{j}^{(e)}=x\right)f_{T_j^{(e)}}(x)dx,    
\end{equation*}
\begin{equation}
=\int_{0}^{\infty}\mathbb{P}\left(T_{i}^{(e)}\leq x\right)f_{T_j^{(e)}}(x)dx=\int_{0}^{\infty}F_{T_{i}^{(e)}}(x)f_{T_j^{(e)}}(x)dx \label{eq:P_excess},   
\end{equation}
where $f_{T_j^{(e)}}(x)$ is defined by Eq.(\ref{eq:Te}). 

Finally, substituting Eq.(\ref{eq:P_excess}) into Eq.(\ref{eq:pi}), and the latter into Eq.(\ref{eq:pdf_N}), we obtain Eq.(\ref{eq:N_final}).
\end{proof}


\subsection{Examples of specific distributions}
\label{sec:specific}

In this section, we look at the distribution of inter-arrival times and the number of packets for four special cases of distributions of inter-generation times at mMTC users: deterministic, uniform, exponential and Pareto distributed.  
The first two are characterized by low variances (increasing failure rates), the third is memoryless, and Pareto distributions are characterized by heavy tails (decreasing failure rates). To ease the presentation, we assume that all mMTC users have i.i.d. traffic generation patterns.  

\subsubsection{Deterministic inter-generation times}
In this case, the pdf of the inter-generation time at the mMTC user is given by
\begin{equation}
f_{\Delta t}(x)=\delta\left(x-\frac{1}{\lambda}\right),    
\label{eq:deterministic}
\end{equation}
where $\frac{1}{\lambda}$ is the (fixed) time between two consecutive generations of data. Substituting Eq.(\ref{eq:deterministic}) into Eq.(\ref{eq:NRx_13}), we get
\begin{eqnarray}
f_{\Delta t^{'}}(x)=\sum_{i=1}^{m}\sum_{k=1}^{N_{max}}\sum_{j=1}^{m}\sum_{l=1}^{N_{max}}p_N(k)p_R(r_i)p_N(l)p_R(r_j)\cdot\nonumber\\
\delta\left(x-\frac{n\Theta}{K}\left(\frac{k}{r_i}-\frac{l}{r_j}\right)-\frac{1}{\lambda}\right),
\label{eq:deterministic_1}    
\end{eqnarray}
and the latter into Eq.(\ref{eq:FTX_aux}), we obtain
\begin{eqnarray}
F_{\Delta t^{'}}(x)=\sum_{i=1}^{m}\sum_{k=1}^{N_{max}}\sum_{j=1}^{m}\sum_{l=1}^{N_{max}}p_N(k)p_R(r_i)p_N(l)p_R(r_j)\nonumber\\
\cdot\int_{0}^{x}\delta\left(w-\frac{n\Theta}{K}\left(\frac{k}{r_i}-\frac{l}{r_j}\right)-\frac{1}{\lambda}\right)dw.
\label{eq:deterministic_2}    
\end{eqnarray}
This yields
\begin{eqnarray}
F_{\Delta t^{'}}(x)=\sum_{i=1}^{m}\sum_{k=1}^{N_{max}}\sum_{j=1}^{m}\sum_{l=1}^{N_{max}}p_N(k)p_R(r_i)p_N(l)p_R(r_j)\nonumber\\
\cdot u\left(x-\frac{n\Theta}{K}\left(\frac{k}{r_i}-\frac{l}{r_j}\right)-\frac{1}{\lambda}\right),
\label{eq:deterministic_2_2}    
\end{eqnarray}
where $u(x)$ denotes the Heaviside function (unit step function)~\cite{Oppenheim}, whose value is $1$ for $x\geq 0$. Otherwise, its value is $0$. In our case, 
as long as $x\geq \frac{n\Theta}{K}\left(\frac{k}{r_i}-\frac{l}{r_j}\right)+\frac{1}{\lambda}$, the value of the $u$-function in Eq.(\ref{eq:deterministic_2_2}) is $1$. Otherwise, it is $0$. 

For mMTC users with homogeneous traffic patterns, Eq.(\ref{eq:theorem_5}) reduces to
\begin{equation}
F_{\Delta t^{'}_{m}}(x)=\lambda^{n}\left(\int_{0}^{x}\left(1-F_{\Delta t^{'}}(w)\right)dw\right)^{n}.
\label{eq:homogeneous_iid}    
\end{equation}

Substituting Eq.(\ref{eq:deterministic_2_2}) into Eq.(\ref{eq:homogeneous_iid}), and computing numerically the integral, we obtain the CDF of the inter-arrival time of data at gNodeB for this case. 

As far as the pmf of the number of packets arriving at gNodeB is concerned, for homogeneous mMTC traffic, Eq.(\ref{eq:N_final}) reduces to
\begin{equation}
p_N(x)=\lambda^{n-1}\sum_{i=1}^{n}p_{N_i}(x)\left(\int_{0}^{\infty}F_{T^{(e)}}(x)\left(1-F_{\Delta t^{'}}(x)\right)dx\right)^{n-1}  
\label{eq:deterministic_final_packets}
\end{equation}
where $f_{T^{(e)}}(x)$ in Eq.(\ref{eq:N_final}) was replaced by Eq.(\ref{eq:Te}), i.e.,
\begin{equation}
f_{T^{(e)}}(x)=\frac{1-F_{\Delta t^{'}}(x)}{\mathbb{E}[\Delta t^{'}]}=\lambda\left(1-F_{\Delta t^{'}}(x)\right).
\label{eq:excess_iid}
\end{equation}
The second term under the sum in Eq.(\ref{eq:deterministic_final_packets}) is not a function of $i$. Hence, it goes out of the sum. Also, as we are dealing with i.i.d. traffic patterns,
\begin{equation}
\sum_{i=1}^{n}p_{N_i}(x)=n p_{N}(x).    
\end{equation}
Now, Eq.(\ref{eq:deterministic_final_packets}) transforms into 
\begin{equation}
p_{N}(x)=n\lambda^{n-1}p_{N}(x)\left(\int_{0}^{\infty}F_{T^{(e)}}(x)\left(1-F_{\Delta t^{'}}(x)\right)dx\right)^{n-1}.
\label{eq:deterministic_final_packets_0}
\end{equation}

In Eq.(\ref{eq:deterministic_final_packets_0}), $F_{\Delta t^{'}}(x)$ is given by Eq.(\ref{eq:deterministic_2_2}), and $F_{T^{(e)}}(x)$ is given by Eq.(\ref{eq:FTe}).


\subsubsection{Uniform inter-generation times}
The pdf of the inter-generation time of data in this case is
\begin{equation}
f_{\Delta t}(x)=\frac{1}{b-a}\left(u(x-a)-u(x-b)\right),
\label{eq:uniform}    
\end{equation}
where $a$ and $b$ are the minimum and maximum possible inter-generation times at the mMTC user, and $u(x)$ is the unit step function introduced previously. 
Note that the average inter-generation time in this case is $\frac{a+b}{2}=\frac{1}{\lambda}$. 

Substituting Eq.(\ref{eq:uniform}) into Eq.(\ref{eq:NRx_13}), and the latter into Eq.(\ref{eq:FTX_aux}), after rearranging we obtain
\begin{align}
F_{\Delta t^{'}}(x)&=\frac{1}{b-a}\sum_{i=1}^{m}\sum_{k=1}^{N_{max}}\sum_{j=1}^{m}\sum_{l=1}^{N_{max}}p_N(k)p_R(r_i)p_N(l)p_R(r_j)\cdot\nonumber\\
&\left(\int_{0}^{x}u\left(w-\frac{n\Theta}{K}\left(\frac{k}{r_i}-\frac{l}{r_j}\right)-a\right)dw\right.\nonumber\\
&\left.-\int_{0}^{x}u\left(w-\frac{n\Theta}{K}\left(\frac{k}{r_i}-\frac{l}{r_j}\right)-b\right)dw\right)
\label{eq:uniform_2}
\end{align}
Finally, replacing Eq.(\ref{eq:uniform_2}) into Eq.(\ref{eq:homogeneous_iid}), and computing numerically the corresponding integrals, we obtain the CDF of the inter-arrival time distribution at gNodeB for this case. 

The pmf of the number of packets arriving at gNodeB, for this case, is obtained from
\begin{equation}
p_{N}(x)=n\lambda^{n-1}p_{N}(x)\left(\int_{0}^{\infty}F_{T^{(e)}}(x)\left(1-F_{\Delta t^{'}}(x)\right)dx\right)^{n-1}
\end{equation}
with $F_{\Delta t^{'}}(x)$ given by Eq.(\ref{eq:uniform_2}), and $F_{T^{(e)}}(x)$ given by Eq.(\ref{eq:FTe}). 

\subsubsection{Exponential inter-generation times}
The pdf of the inter-generation time of data in this case is 
\begin{equation}
f_{\Delta t}(x)=\lambda e^{-\lambda x},    
\label{eq:exponential_0}
\end{equation}
where $\frac{1}{\lambda}$ is the average inter-generation time. Substituting Eq.(\ref{eq:exponential_0}) into Eq.(\ref{eq:NRx_13}), after rearranging we obtain
\begin{align}
f_{\Delta t^{'}}(x)&=\lambda e^{-\lambda x}\sum_{i=1}^{m}\sum_{k=1}^{N_{max}}\sum_{j=1}^{m}\sum_{l=1}^{N_{max}}p_N(k)p_R(r_i)p_N(l)p_R(r_j)\nonumber\\
&\cdot e^{\frac{\lambda n K}{\Theta}\left(\frac{k}{r_i}-\frac{l}{r_j}\right)}.
\label{eq:exponential}
\end{align}
Replacing Eq.(\ref{eq:exponential}) into Eq.(\ref{eq:FTX_aux}), after solving the corresponding integral and rearranging, we obtain
\vspace{-6pt}

\footnotesize
\begin{align}
F_{\Delta t^{'}}(w)&=\left(1-e^{-\lambda w}\right)\sum_{i=1}^{m}\sum_{k=1}^{N_{max}}\sum_{j=1}^{m}\sum_{l=1}^{N_{max}}p_N(k)p_R(r_i)p_N(l)p_R(r_j)\nonumber\\
&\cdot e^{\frac{\lambda n K}{\Theta}\left(\frac{k}{r_i}-\frac{l}{r_j}\right)}.
\label{eq:exponential_2}
\end{align}
\normalsize
Finally, substituting Eq.(\ref{eq:exponential_2}) into Eq.(\ref{eq:homogeneous_iid}), solving the corresponding integral, and rearranging, we get
\begin{equation}
F_{\Delta t_{m}}^{'}(x)=\lambda^{n} \left[(1-A)x+\frac{A}{\lambda}\left(1-e^{-\lambda x}\right)\right]^{n},
\label{eq:exponential_3}    
\end{equation}
where $$A=\sum_{i=1}^{m}\sum_{k=1}^{N_{max}}\sum_{j=1}^{m}\sum_{l=1}^{N_{max}}p_N(k)p_R(r_i)p_N(l)p_R(r_j)e^{\frac{\lambda n K}{\Theta}\left(\frac{k}{r_i}-\frac{l}{r_j}\right)}$$
As can be observed from Eq.(\ref{eq:exponential_3}), in this scenario we can obtain a closed-form expression for the CDF of inter-arrival time at  gNodeB. 

The pmf of the number of packets arriving at gNodeB is
\begin{equation}
p_{N}(x)=n\lambda^{n-1}p_{N}(x)\left(\int_{0}^{\infty}F_{T^{(e)}}(x)\left(1-F_{\Delta t^{'}}(x)\right)dx\right)^{n-1}
\end{equation}
with $F_{\Delta t^{'}}(x)$ given by Eq.(\ref{eq:exponential_2}), and $F_{T^{(e)}}(x)$ given by Eq.(\ref{eq:FTe}). 

\subsubsection{Pareto-distributed inter-generation times}
The pdf of the inter-generation time in this case is
\begin{equation}
f_{\Delta t}(x)=\frac{\alpha a^{\alpha}}{x^{\alpha+1}},
\label{eq:Pareto_0}    
\end{equation}
where $a$ is the minimum possible inter-generation time, and $\alpha$ is the shape parameter of the Pareto distribution. We assume that $\alpha>1$. The average of Pareto distribution is $\frac{1}{\lambda}=\frac{a\alpha}{\alpha-1}$, resulting in $a=\frac{\alpha-1}{\alpha\lambda}$.

The pdf of the inter-arrival time at gNodeB is (after replacing Eq.(\ref{eq:Pareto_0}) into Eq.(\ref{eq:NRx_13}))
\begin{align}
f_{\Delta t^{'}}(x)&=\sum_{i=1}^{m}\sum_{k=1}^{N_{max}}\sum_{j=1}^{m}\sum_{l=1}^{N_{max}}p_N(k)p_R(r_i)p_N(l)p_R(r_j)\nonumber\\
&\cdot\frac{\alpha a^\alpha}{\left(x-\frac{n\Theta}{K}\left(\frac{k}{r_i}-\frac{l}{r_j}\right)\right)^{\alpha+1}}.
\label{eq:Pareto_2}    
\end{align}
Substituting Eq.(\ref{eq:Pareto_2}) into Eq.(\ref{eq:FTX_aux}), after solving the integral and rearranging, we obtain
\begin{eqnarray}
F_{\Delta t^{'}}(x)=a^{\alpha}\sum_{i=1}^{m}\sum_{k=1}^{N_{max}}\sum_{j=1}^{m}\sum_{l=1}^{N_{max}}p_N(k)p_R(r_i)p_N(l)p_R(r_j)\nonumber\\
\cdot\left(\hspace{-3pt}-\hspace{-3pt}\left(x\hspace{-3pt}-\hspace{-3pt}\frac{n\Theta}{K}\hspace{-3pt}\left(\frac{k}{r_i}\hspace{-3pt}-\hspace{-3pt}\frac{l}{r_j}\right)\right)^{\hspace{-3pt}-\alpha}\hspace{-3pt}\hspace{-3pt}+\hspace{-3pt}\left(\hspace{-3pt}-\hspace{-3pt}\frac{n\Theta}{K}\hspace{-3pt}\left(\frac{k}{r_i}\hspace{-3pt}-\hspace{-3pt}\frac{l}{r_j}\right)\right)^{\hspace{-3pt}-\alpha}\right)
\label{eq:Pareto}
\end{eqnarray}
Finally, substituting Eq.(\ref{eq:Pareto}) into Eq.(\ref{eq:homogeneous_iid}), solving the integral and rearranging the terms, we obtain the CDF of the inter-arrival time of data packets at gNodeB as Eq.(\ref{eq:Pareto_3}), where $B=\frac{n\Theta}{K}\left(\frac{k}{r_i}-\frac{l}{r_j}\right)$. In this scenario as well, we can find the distribution of the inter-arrival time at gNodeB in closed form.  
\begin{figure*}[b]
\begin{equation}
F_{\Delta t_{m}^{'}}(x)=\lambda^{n}\left[x+a^{\alpha}\sum_{i=1}^{m}\sum_{k=1}^{N_{max}}\sum_{j=1}^{m}\sum_{l=1}^{N_{max}}p_N(k)p_R(r_i)p_N(l)p_R(r_j)\left(\frac{\left(x-B\right)^{1-\alpha}}{1-\alpha}-\frac{\left(-B\right)^{1-\alpha}}{1-\alpha}-\left(-B\right)^{-\alpha}x\right)\right]^{n}
\label{eq:Pareto_3}
\end{equation}
\end{figure*}

The pmf of the number of packets arriving at gNodeB is
\begin{equation}
p_{N}(x)=n\lambda^{n-1}p_{N}(x)\left(\int_{0}^{\infty}F_{T^{(e)}}(x)\left(1-F_{\Delta t^{'}}(x)\right)dx\right)^{n-1},
\end{equation}
with $F_{\Delta t^{'}}(x)$ given by Eq.(\ref{eq:Pareto}), and $F_{T^{(e)}}(x)$ given by Eq.(\ref{eq:FTe}). 

\section{Performance Evaluation}
\label{sec:sims}
We first describe the simulation setup. Then, we focus on the impact some of the parameters have on the base station traffic patterns. 

\subsection{Simulation setup}
As input parameters for the signal quality, we have used a 5G trace with data measured in the Republic of Ireland. 
These traces can be found in~\cite{trace_5G_link}, with a detailed description in~\cite{trace_paper_5G}.
The parameter of interest from the trace is CQI (Channel Quality Indicator) with $15$ levels, which serves to determine the per-block rate of a user in a frame. These measurements were conducted for one user, but on different days, for different applications, and when the user is static and driving  around. To mimic the predominantly static nature of mMTC devices, we picked 6 ``basic''\footnote{We call them basic users as we will replicate these users to create more users in the corresponding scenarios, as needed.} users that are static, and assume they all are in the same cell. 
Based on the frequency of occurrence of a per-block rate for every user, we obtained the corresponding per-block rate probabilities in Table~\ref{Table-R_2}.   

The packet size in all the scenarios is $5$\:kbits.\footnote{We tried other values for  packet sizes with similar conclusions drawn.} Unless stated otherwise, mMTC users are placed uniformly in a distance $(0.1,3)$\:km from the gNodeB. The propagation delay for every user is found as $t_{prop}=\frac{d}{c}$, where $d$ is the distance of the user from gNodeB, and $c=3\cdot10^8$\:m/s.  

\textcolor{black}{The development of network slicing in 5G~\cite{p11} has allowed operators to split users into groups, with users of similar use cases, like mMTC users in the same cell, 
within the same group (slice). In the following, all the mMTC users belong to the same slice.}

The frame duration is $10$\:ms.  \textcolor{black}{The subcarrier spacing is $30$\:KHz, with $12$ subcarriers per block, making the block width $360$\:KHz.} 
\textcolor{black}{The total number of PRBs is $275$~\cite{3GPP_5G_NR}}. 

The simulations are conducted in MATLAB R2021a. We take the average of the metrics of interest over $10000$ runs.
\begin{table*}[t]
\caption{Per-block rates and the corresponding probabilities for users from the Republic of Ireland trace~\cite{trace_paper_5G}}
\label{Table-R_2}
\centering
\begin{tabular}{|c|c|c|c|c|c|c|c|c|c|c|c|c|c|c|c|}
\hline
R (kbps) & 48 & 73.6 & 121.8 & 192.2 & 282 & 378 & 474.2 & 712 & 772.2 & 874.8 & 1063.8 & 1249.6 & 1448.4 & 1640.6 & 1778.4\\
\hline
$p_{1,k}$ & 0 & 0 & 0 & 0 & 0 & 0 & 0.01 & 0.05 & 0.11 & 0.13 & 0.14 & 0.18 & 0.06 & 0.11 & 0.21\\
\hline   
$p_{2,k}$ & 0 & 0 & 0 & 0 & 0 & 0.01 & 0.02 & 0.06 & 0.13 & 0.14 & 0.2 & 0.21 & 0.07 & 0.09 & 0.07\\
\hline
$p_{3,k}$ & 0.01 & 0 & 0 & 0 & 0 & 0.01 & 0.01 & 0.02 & 0.06 & 0.13 & 0.17 & 0.18 & 0.08 & 0.18 & 0.15\\
\hline
$p_{4,k}$ & 0 & 0 & 0 & 0 & 0 & 0.02 & 0.03 & 0.13 & 0.06 & 0.2 & 0.32 & 0.11 & 0.01 & 0.09 & 0.03\\
\hline                                                         
$p_{5,k}$ & 0 & 0 & 0 & 0 &  0 & 0 & 0.04 & 0.07 & 0.13 & 0.17 & 0.22 & 0.2 & 0.05 & 0.06 & 0.06\\
\hline 
$p_{6,k}$ & 0 & 0 & 0 & 0 & 0.01 & 0.03 & 0.11 & 0.12 & 0.19 & 0.15 & 0.15 & 0.12 & 0.05 & 0.04 & 0.03\\
\hline                                                         
\end{tabular}
\end{table*}
\begin{figure*}[t]
\begin{minipage}{0.33\linewidth}
\centering
\includegraphics[width=\textwidth]{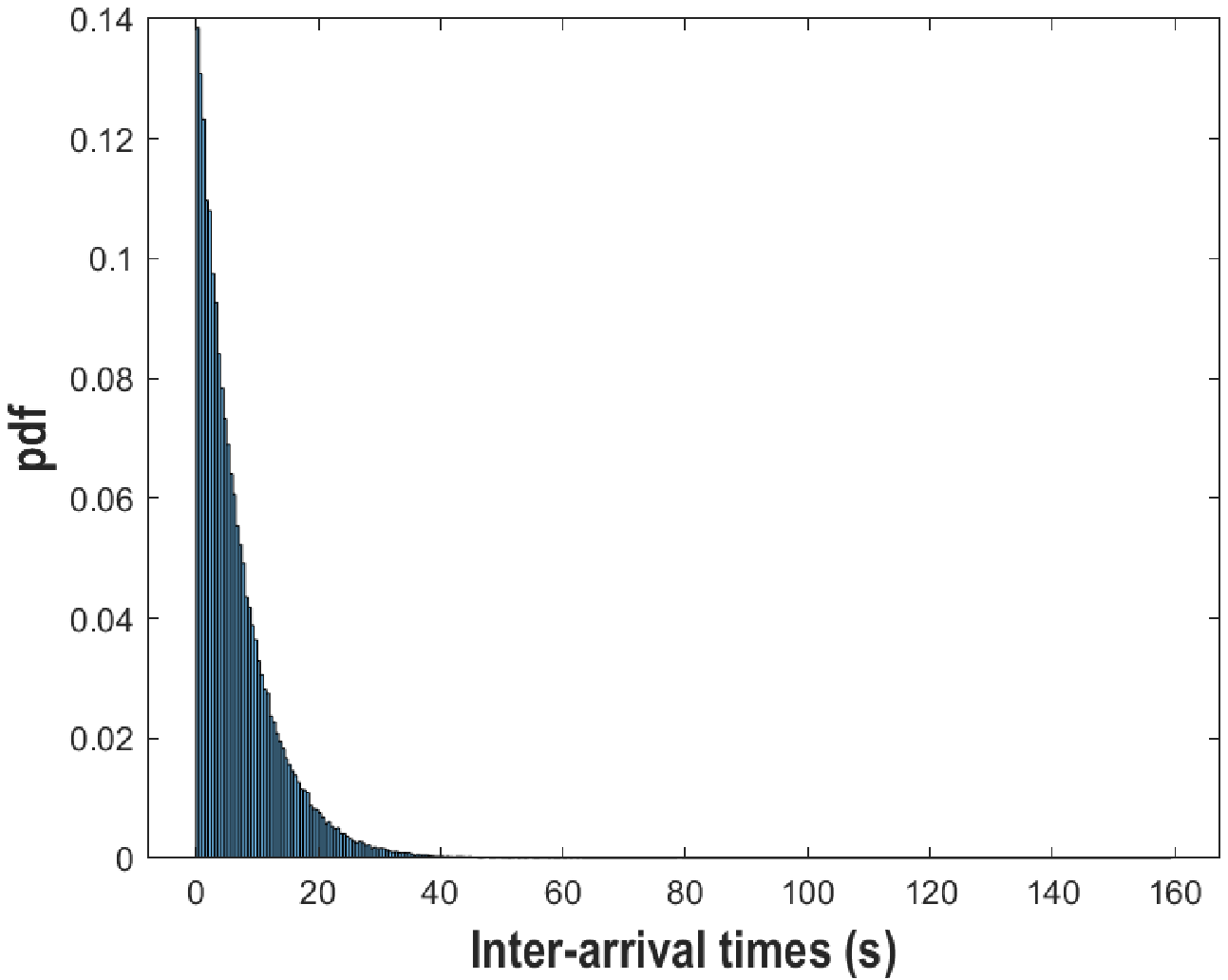}
\caption{The probability density function of inter-arrival times for uniform inter-generation times.}
\label{fig:pdf1}
\end{minipage}
\begin{minipage}{0.33\linewidth}
\centering
\includegraphics[width=\textwidth]{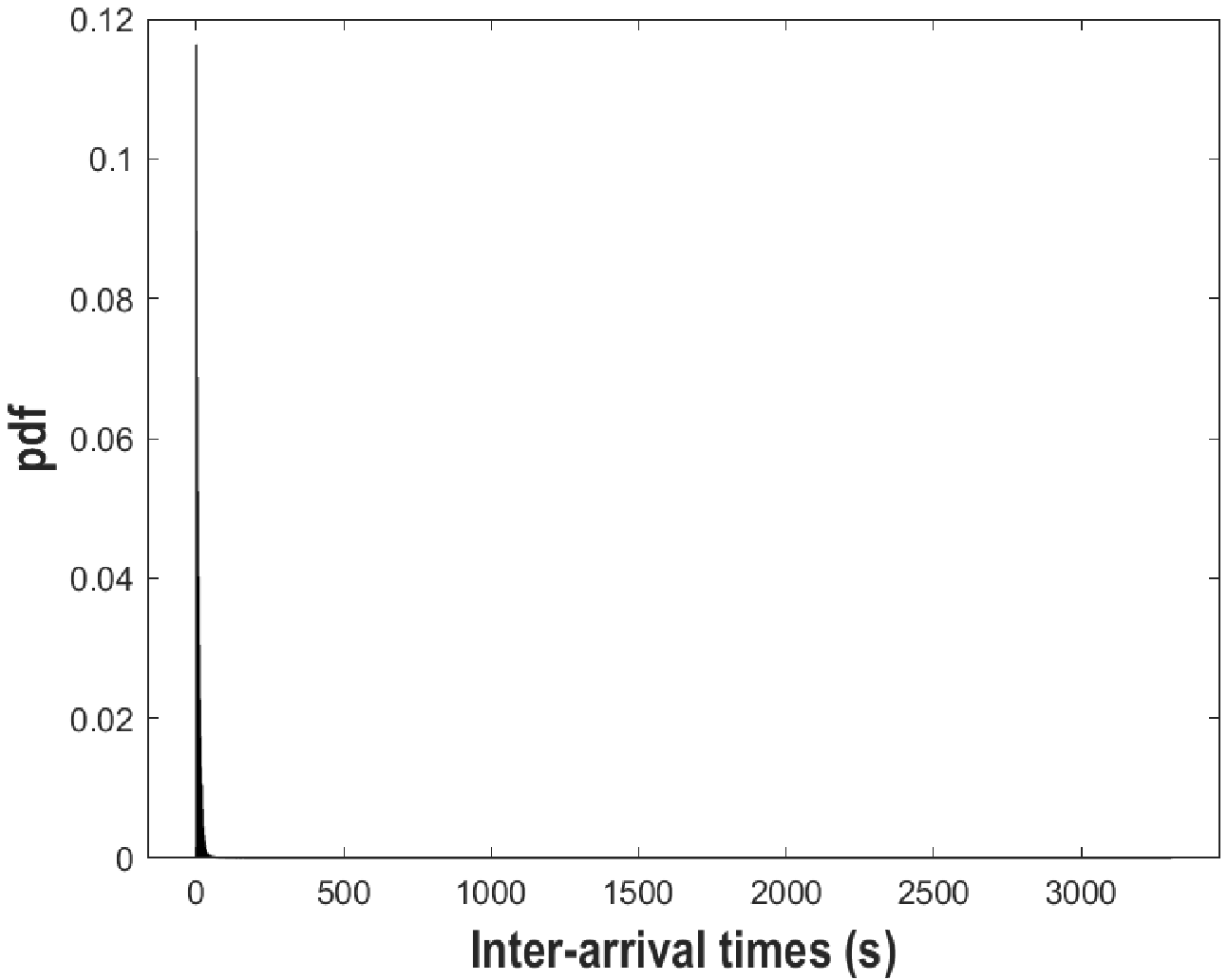}
\caption{The probability density function of inter-arrival times for B. Pareto inter-generation times.}
\label{fig:pdf2}
\end{minipage}
\begin{minipage}{0.33\linewidth}
\centering
\includegraphics[width=\textwidth]{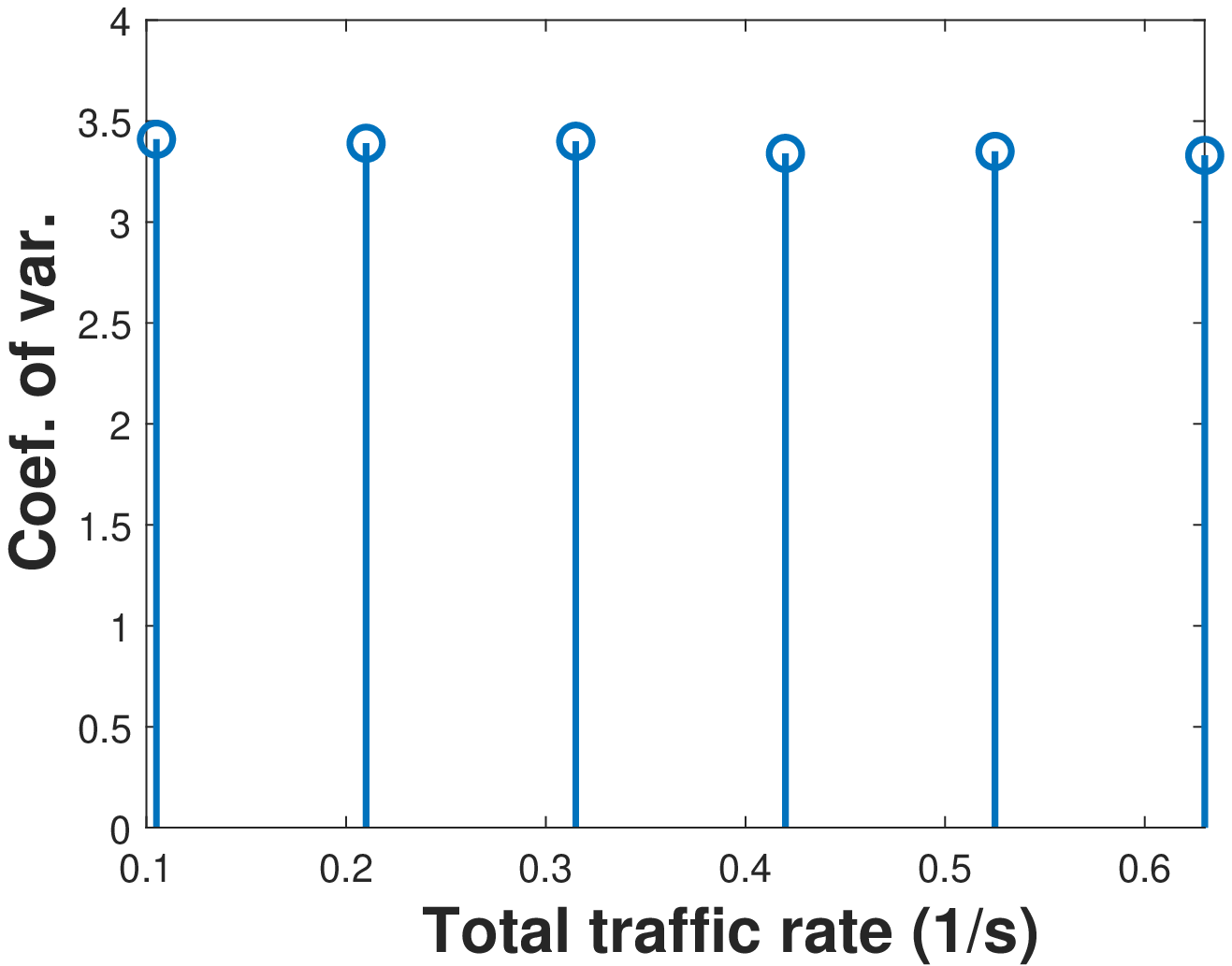}
\caption{Traffic variability at gNodeB vs. generation rate.}
\label{fig:cv_lambda}
\end{minipage}
\end{figure*}

\subsection{Impact of channel characteristics}
First, we consider the impact of channel conditions of mMTC users on the traffic pattern at gNodeB. We do this for homogeneous users, i.e., in a given scenario all mMTC users have i.i.d. traffic generation patterns. In the first case, users have uniform inter-generation times, whereas in the second their inter-generation times are drawn from a Bounded-Pareto distribution. 

\subsubsection{Inter-arrival times for uniformly distributed inter-generation times}
In the first scenario, the inter-generation times are uniform, and the users of interest belong to type-1, type-3, and type-5 users (Table~\ref{Table-R_2}). The total number of users is $n=1500$. The number of generated packets at once is drawn uniformly from the interval $(10,20)$. 
We consider the outcomes for three different inter-generation rates: $\lambda_1=0.01 s^{-1}$, $\lambda_2=0.05 s^{-1}$, and $\lambda_{3}=0.1 s^{-1}$. Table~\ref{Table-uniform} shows the results for the first, second, and third moments of the inter-arrival times at gNodeB in all the scenarios. Table~\ref{Table-uniform} shows also the coefficient of variation of the inter-arrival times at gNodeB, defined as
\begin{equation}
c_v(X)=\frac{\sqrt{\text{Var}(X)}}{\mathbb{E}[X]},   
\end{equation}
which serves as the most suitable quantification for the variability of a stochastic process. 
As can be observed from Table~\ref{Table-uniform}, the statistics for the same inter-generation rate are almost identical for all three user types. This implies that the \emph{channel statistics of the mMTC users have almost no impact on the traffic pattern at gNodeB}. The second thing to observe is that increasing $\lambda$ decreases the inter-arrival time of the traffic at gNodeB. This is expected as decreasing the time between generation instants will decrease the inter-arrival time as well.


\subsubsection{Inter-arrival times for Bounded Pareto inter-generation times}
\textcolor{black}{In the second set of results, the inter-generation times are Bounded-Pareto, characterized by the shape parameter $\alpha$, and minimum and maximum values $L$ and $H$, respectively. In all cases that follow, $\alpha=1.95$, whereas $L$ and $H$ are adjusted to achieve a given average inter-generation time. \textcolor{black}{When $\lambda=0.01 s^{-1}$, $L=48.75$ and $H=10000$. For $\lambda=0.05 s^{-1}$, $L=9.81$ and $H=2000$. Finally, for $\lambda=0.1 s^{-1}$, $L=4.905$ and $H=1000$.}}

\textcolor{black}{Table~\ref{Table-Pareto} shows the results for the same cases corresponding to those of Table~\ref{Table-uniform}, but for different user types (2, 4, and 6 now). Similar conclusions apply regarding the lack of impact of channel characteristics on the network traffic pattern. The variability is higher in this case (the value of the coefficient of variation can go up to $4.83$). The other thing to observe in this case as well is that \emph{increasing the traffic generation rates does not increase the coefficient of variation of inter-arrival times at gNodeB for any mMTC user type.}}

Fig.~\ref{fig:pdf1} illustrates the probability density function of inter-arrival times at gNodeB for user-type 1 and $\lambda=0.01 s^{-1},$ when inter-generation time is uniform, whereas Fig.~\ref{fig:pdf2} depicts the pdf for Bounded-Pareto inter-generation times for the same user type and the same average inter-generation time. Comparing Figs.~\ref{fig:pdf1} and~\ref{fig:pdf2}, we can observe that in the second case there are instances when the inter-arrival time can be $20\times$ higher than the maximum inter-arrival time in the first case. This is due to the high variance of Bounded Pareto distribution, classifying it as a  heavy-tailed distribution. 
\noindent\emph{Note:} In Figs.~\ref{fig:pdf1} and~\ref{fig:pdf2}, for better visibility of the graphs, we have shown the results for $15$ mMTC users only.  

The fact that the traffic pattern is \emph{almost completely insensitive} on the channel conditions of mMTC users has an important practical implication. Namely, network planning can be conducted based solely on the number of users and their traffic, but not on their channel characteristics. The reason lies in the nature of mMTC traffic, in which packets are small and traffic is relatively sparse, making inter-arrival times non-dependent on the per-block rate distribution. 

\begin{table}[t]
\caption{The first three moments and the coefficient of variation for user types 1, 3, and 5, for three uniform inter-generation times}
\label{Table-uniform}
\centering
\begin{tabular}{|c|c|c|c|c|}
\hline
$\lambda_1=0.01 s^{-1}$ & $\mathbb{E}[X]$ & $\mathbb{E}[X^2]$ & $\mathbb{E}[X^3]$ & $c_V$ \\
\hline
user 1 & 0.068 & 0.037 & 2.95 & 2.59\\
\hline   
user 3 & 0.068 & 0.037 & 2.95 & 2.59 \\
\hline
user 5 & 0.068 & 0.037 & 2.95 & 2.59 \\
\hline
\hline
$\lambda_3=0.05 s^{-1}$ & $\mathbb{E}[X]$ & $\mathbb{E}[X^2]$ & $\mathbb{E}[X^3]$ & $c_V$ \\
\hline                                                         
user 1 & 0.0136 & 0.0015 & 0.0236 & 2.59 \\
\hline 
user 3 & 0.0136 & 0.0015 & 0.0236 & 2.59 \\
\hline                   
user 5 & 0.0136 & 0.0015 & 0.0235 & 2.59 \\
\hline
\hline
$\lambda_5=0.1 s^{-1}$ & $\mathbb{E}[X]$ & $\mathbb{E}[X^2]$ & $\mathbb{E}[X^3]$ & $c_V$ \\
\hline                                                         
user 1 & 0.0068 & $3.72\cdot 10^{-4}$ & 0.003 & 2.6 \\
\hline 
user 3 & 0.0068 & $3.71\cdot10^{-4}$ & 0.003 & 2.59 \\
\hline                   
user 5 & 0.0068 & $3.71\cdot 10^{-4}$ & 0.003 & 2.59 \\
\hline
\end{tabular}
\end{table}
\begin{table}[t]
\caption{The first three moments and the coefficient of variation for user types 2, 4, and 6, for three Bounded Pareto inter-generation times}
\label{Table-Pareto}
\centering
\begin{tabular}{|c|c|c|c|c|}
\hline
$\lambda_1=0.01 s^{-1}$ & $\mathbb{E}[X]$ & $\mathbb{E}[X^2]$ & $\mathbb{E}[X^3]$ & $c_V$ \\
\hline
user 2 & 0.0697 & 0.156 & 253.45 & 4.83\\
\hline   
user 4 & 0.0697 & 0.156 & 253.29 & 4.83 \\
\hline
user 6 & 0.0697 & 0.156 & 253.28 & 4.83 \\
\hline
\hline
$\lambda_3=0.05 s^{-1}$ & $\mathbb{E}[X]$ & $\mathbb{E}[X^2]$ & $\mathbb{E}[X^3]$ & $c_V$ \\
\hline                                                         
user 2 & 0.014 & 0.0063 & 2.02 & 4.83 \\
\hline 
user 4 & 0.014 & 0.0062 & 2.04 & 4.82 \\
\hline                   
user 6 & 0.014 & 0.0062 & 2.04 & 4.82 \\
\hline
\hline
$\lambda_5=0.1 s^{-1}$ & $\mathbb{E}[X]$ & $\mathbb{E}[X^2]$ & $\mathbb{E}[X^3]$ & $c_V$ \\
\hline                                                         
user 2 & 0.007 & 0.0016 & 0.25 & 4.83 \\
\hline 
user 4 & 0.007 & 0.0016 & 0.24 & 4.83 \\
\hline                   
user 6 & 0.007 & 0.0016 & 0.25 & 4.83 \\
\hline
\end{tabular}
\end{table}

\subsection{Impact of generation rates on gNodeB traffic variability}
Next, we look at how the inter-generation rates ($\lambda$) affect the variability of gNodeB traffic. We consider heterogeneous users now.
Table~\ref{traffic} depicts the traffic characteristics of the mMTC users. In total, there are 1500 mMTC users in the cell. 
The number of packets generated at once by mMTC is uniform with the values $N_i$ from the corresponding intervals in Table~\ref{traffic}. We look at the coefficient of variation of inter-arrival times at gNodeB. 

Fig.~\ref{fig:cv_lambda} illustrates the results. The values on the x-axis correspond to the sum of $\lambda$ over all six types for single users. The first value on the x-axis is $0.01+0.02+0.03+0.011+0.019+0.015=0.0905 s^{-1}$. In the second case, we multiply by $2$ the generation rates of every user, hence $2\times$ higher value of the x-axis. This is repeated $6\times$ (hence $6$ values in the plot of Fig.~\ref{fig:cv_lambda}). So, the highest value on the x-axis corresponds to traffic generation rates that are $6\times$ higher for all the users than those corresponding to the basic scenario from Table~\ref{traffic}.  


The interesting outcome from Fig.~\ref{fig:cv_lambda} is that the variability of gNodeB traffic is oblivious to the traffic generation rates of mMTC users ($c_V$ remains almost unchanged at all with the increase in traffic generation rates). 
\begin{table}[t]
\caption{Basic scenario with heterogeneous mMTC users}
\label{traffic}
\centering
\begin{tabular}{|c|c|c|c|c|}
\hline
Users & $\lambda_i$ & $N_i$ & inter-gen. time dist. & number of users\\
\hline
type 1 & 0.01 & (10,20) & uniform & 250\\
\hline   
type 2 & 0.02 & (10,15) & deterministic & 250 \\
\hline
type 3 & 0.03 & (10,25) & exponential & 250 \\
\hline
type 4 & 0.011 & (10,20) & deterministic & 250\\
\hline   
type 5 & 0.019  & (10,15) & exponential & 250 \\
\hline
type 6 & 0.015  & (10,15) & uniform & 250\\
\hline  
\end{tabular}
\end{table}

\subsection{Impact of number of users on gNodeB traffic variability}
Having shown the (lack of) impact of traffic generation rates on the variability of network traffic, we look next at how increasing the number of users affects the coefficient of variation of inter-arrival times at gNodeB. We consider the same basic scenario from Table~\ref{traffic}, with one exception. The number of users changes from $50$ to $300$ for each user-types, i.e., the total number of users ranges from $300$ to $1800$. Fig.~\ref{fig:cv_n} shows the coefficient of variation for different number of users. As can be observed from Fig.~\ref{fig:cv_n}, increasing the number of users increases the coefficient of variation of inter-arrival times \textcolor{black}{almost linearly}.   

\subsection{Impact of mMTC distance on traffic characteristics}
Having looked at the impact the generation rate and the number of mMTC users have on network traffic variability, we proceed with investigating the impact of distance of mMTC from the base station. The other data are from Table~\ref{traffic}. We consider four scenarios to that end. In Scenario~1, each mMTC user is placed uniformly in the range $(0.1,0.5)$\:km. In Scenario~2, the range where mMTC are is $(0.1,1)$\:km (again uniform), whereas in Scenario~3 every device is placed uniformly in the range $(0.1,2)$\:km. Finally, in Scenario~4, the position of each mMTC user is drawn uniformly from the range $(0.1,3)$\:km. Needless to say, but mMTC users' positions are fixed.  

\begin{figure*}[t]
\begin{minipage}{0.33\linewidth}
\centering
\includegraphics[width=\textwidth]{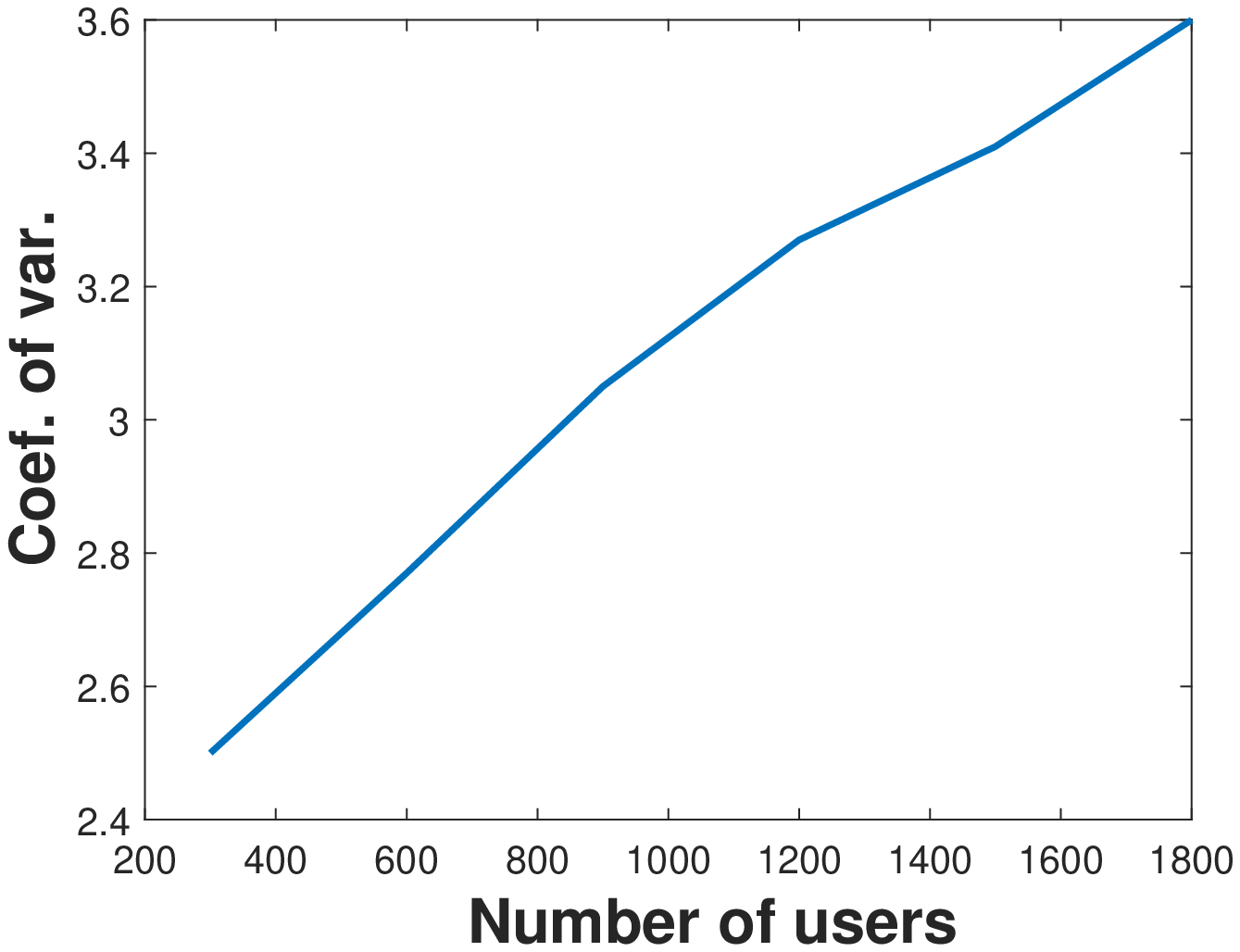}
\caption{Traffic variability at gNodeB vs. number of users.}
\label{fig:cv_n}
\end{minipage}
\begin{minipage}{0.33\linewidth}
\centering
\includegraphics[width=\textwidth]{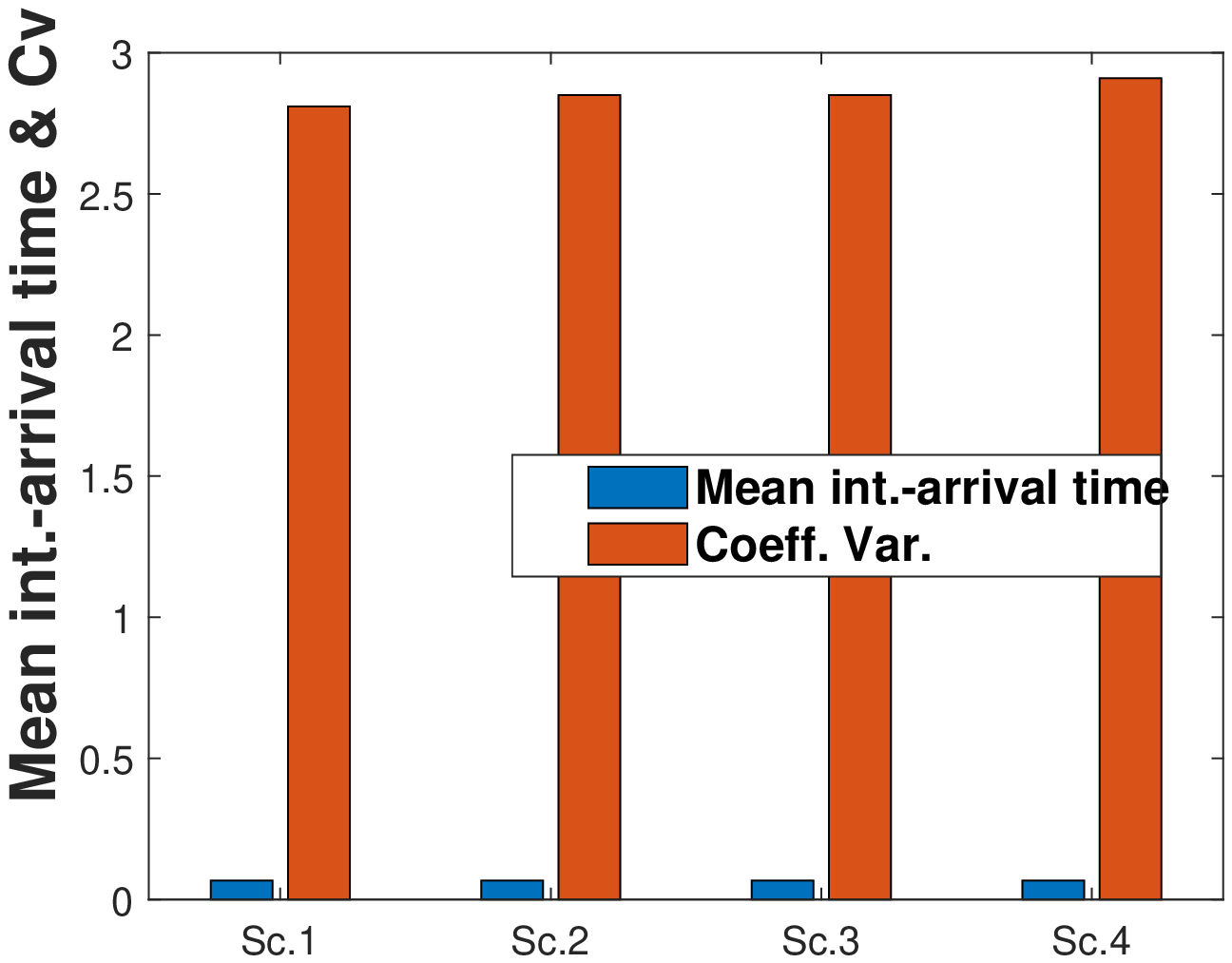}
\caption{Traffic statistics at gNodeB for different distances.}
\label{fig:cv_d}
\end{minipage}
\begin{minipage}{0.33\linewidth}
\centering
\includegraphics[width=\textwidth]{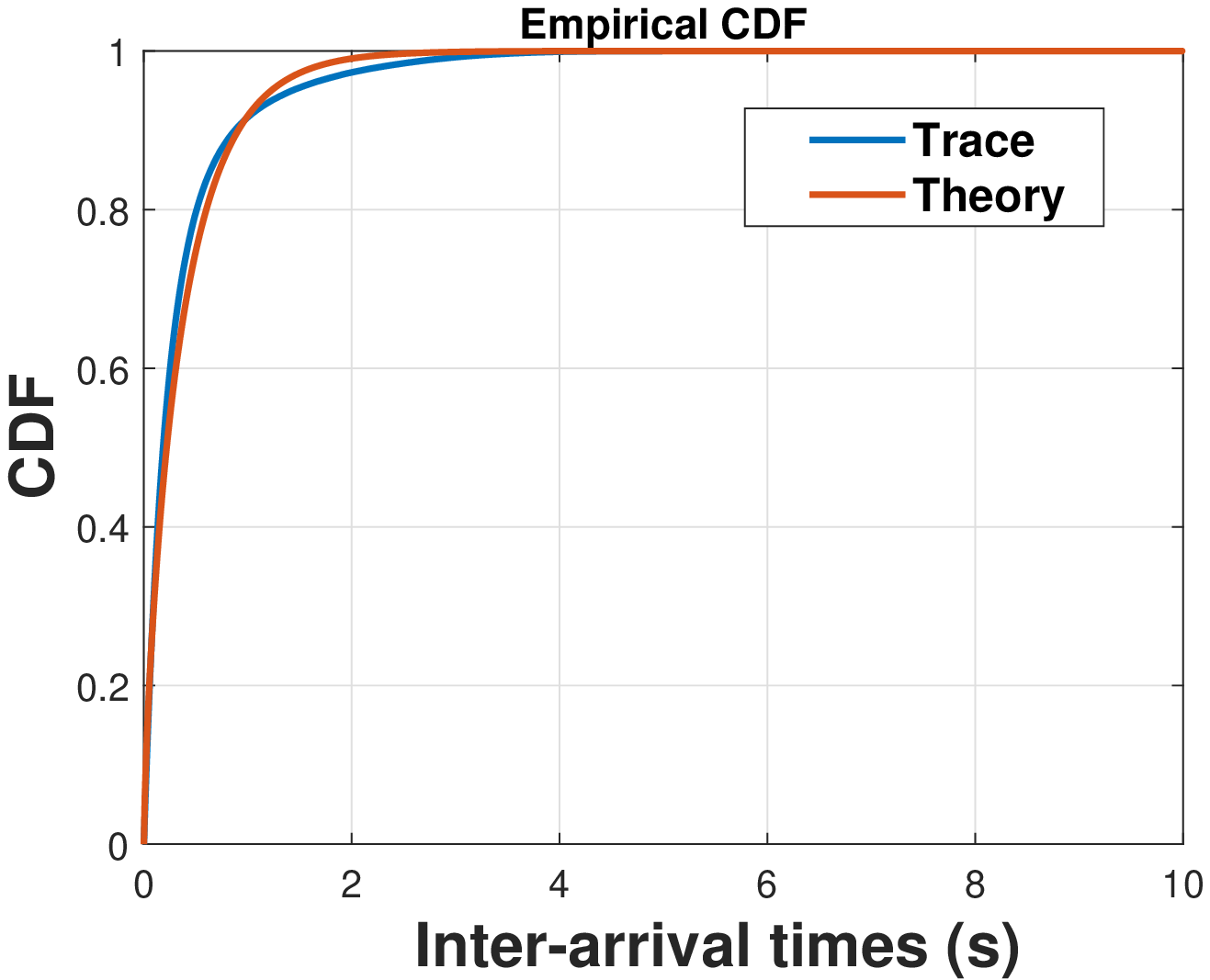}
\caption{The CDF of the inter-arrival times from the trace and theory.}
\label{fig:cdf_trace}
\end{minipage}
\end{figure*}

Fig.~\ref{fig:cv_d} depicts the values of the average inter-arrival times and their $c_V$ for the four scenarios. The impact of the distance is almost invisible on the gNodeB traffic characteristics. The reason is that the propagation delay is 
very low and fixed for every device (mMTC position is fixed), making very little or no impact at all on the network traffic pattern.   

\textcolor{black}{\subsection{Validation on a dataset}
In the last part of this section, we validate our theoretical result for the inter-arrival time at gNodeB. We use a trace of data from $21$ IoT devices as input parameters~\cite{IoT_trace}. More explanations on this dataset can be found in~\cite{IoT_trace_paper}. For every device, we randomly choose a user type and then the corresponding per-block rate from the other trace (Table~\ref{Table-R_2}). From the dataset~\cite{IoT_trace}, it can be observed that data transmission follows different patterns at different devices. In some of these, data are sent uniformly several times during $1$\:s, whereas in others every $5$\:s. There are no data available on the number of transmitted packets per time. We assume that 30 packets are transmitted at once. The other parameters are as in the previous scenarios. All the users are within the same cell. We observed that the coefficient of variation of inter-generation times for all the users is in the range $(0.5,0.6)$. Fig.~\ref{fig:cdf_trace} shows the CDF of inter-arrival times at gNodeB from these users. We model the behavior of every user in the system, to the best possible extent, by the corresponding distribution of inter-generation times (the number of packets is fixed whenever there is a transmission). Then, we use Eq.(\ref{eq:theorem_5}) to derive the CDF of the inter-arrival time at gNodeB. Fig.~\ref{fig:cdf_trace} shows this curve as well. As can be observed from Fig.~\ref{fig:cdf_trace}, our model can predict quite accurately the inter-arrival time distribution in this realistic scenario, despite the fact that the model may not perfectly capture all the intricacies of the generation process on the users' side.}           

\section{Related Work}
\label{sec:related}
While the traffic a single user generates, be it an enhanced Mobile Broadband (eMBB), URLLC, or mMTC, can be determined relatively easily depending on the application/service, predicting the traffic pattern (inter-arrival time and size distribution) at the base station from a (usually) large number of mMTC devices is not straightforward. To our best knowledge, there are no other works that can predict analytically the exact distribution of the traffic pattern at the base station.     
Regarding the traffic distribution at single mMTC users, the general assumption is that inter-generation times for sparse traffic are either periodic (i.e., deterministic)~\cite{p14}, uniform~\cite{Hao19},~\cite{Alhussien20}, or exponential~\cite{Ortiz} for sparse traffic, and for mMTC users with more intense traffic, the number of packets underlie a beta distribution~\cite{Wang2020}, whereas the inter-generation time is governed by a Pareto distribution~\cite{Hao19}. In all these models, the sizes of the packets are fixed, and very small. However, in those works, there are no indications as to what distribution the traffic arriving from a large number of mMTC devices to the base station might have. On the other hand, in our paper, we assume a general distribution for the inter-generation times as well as for the number of packets generated at a single mMTC unit, and derive the distribution of the traffic pattern at gNodeB. Furthermore, for most of these single-mMTC traffic distributions, 
we derive the traffic distribution at gNodeB. We consider as special cases the inter-generation times being deterministic, uniform, exponential, and Pareto distributed. While one of these works~\cite{Wang2020} assumes the number of generated packets underlies a beta distribution, our approach is more general, as the number of packets underlies any generic distribution, and it can be easily adjusted to capture this special case.   

Predicting certain parameters of the traffic arriving at the base station using machine learning is done in~\cite{Weerasinghe19} and~\cite{Soraa20}. In~\cite{Weerasinghe19}, the authors propose a supervised learning approach that predicts the number of arrivals and when a burst of data will arrive for the number of packets that follow a beta distribution. However, the service type of interest in~\cite{Weerasinghe19} is URLLC. On the other hand, in our paper, we derive the exact distribution of both the number of packets arriving at the base station and the inter-arrival time between two data batches for general traffic generations at mMTC users. The main contribution of~\cite{Soraa20} is proposing a machine learning-based model for predicting possible occurrences of congestion under bursty traffic conditions, where the latter is modeled with a beta distribution. Contrary to~\cite{Soraa20}, our work is more general because deriving the exact distribution of the traffic pattern at gNodeB can also predict whether network congestion is expected to occur for any type of mMTC data distributions.     

\section{Conclusion}
\label{sec:conclusion}
In this paper, we have derived the inter-arrival time distribution of the traffic from different mMTC users within the cell to the base station. We assumed a generic traffic pattern at each mMTC user, specified by a general inter-generation time distribution and a general number of packets transmitted at once. 
\textcolor{black}{Conducting simulations on a real-life trace, results show that 
the variability of the traffic pattern increases 
when increasing the number of mMTC users, but  
it is not affected by the increase in traffic generation rates.} 
In our future work, we plan to characterize the traffic pattern of the other two service types in 5G - URLLC and eMBB. 

\bibliographystyle{ieeetr}
\bibliography{Fidan}

\end{document}